\documentclass[10pt,a4paper]{article}
\usepackage[utf8]{inputenc}
\usepackage{amsmath}
\usepackage{amsfonts}
\usepackage{xcolor, soul}

\sethlcolor{yellow}
\usepackage{lipsum}
\usepackage{amssymb}
\usepackage{graphicx}
\usepackage{multirow}
\usepackage[many]{tcolorbox}
\usepackage[left=1.5cm,right=1.5cm,top=1.5cm,bottom=1.5cm]{geometry}
\usepackage{tikz}
\usepackage{pgfplots}
\usepackage{caption}
\usepackage{subcaption}
\pgfplotsset{width=10cm,compat=1.9}
\usepackage{graphicx}
\usepackage{hyperref}
\DeclareUnicodeCharacter{2212}{-}
\usepackage{cite}

\title{Inflection Point of Minimally Coupled Tachyonic Scalar Field}

\author{Jaskirat Kaur\footnote{jaskiratkaur653@gmail.com}, \quad   S.D.Pathak\footnote{prince.pathak19@gmail.com}\\ Department of Physics, Lovely Professional University,\\ Phagwara, 144411, Punjab, India\\
Maxim Khlopov
\footnote{khlopov@apc.in2p3.fr}\\
Virtual Institute of Astroparticle Physics, 75018 Paris, France\\
Institute of Physics, Southern Federal University, Rostov on Don 344090, Russia\\ National Research Nuclear University MEPHI, 115409 Moscow, Russia\\
Manabendra Sharma\footnote{sharma.man@mahidol.ac.th}\\
Centre for Theoretical Physics and Natural Philosophy, \\Nakhonsawan Studiorum for Advanced Studies, Mahidol University, Nakhonsawan, 60130, Thailand}
\date{}

\begin{document}
\maketitle

\begin{abstract}
In this paper, we investigate the behaviour of a minimally coupled tachyonic scalar field at inflection point in an accelerating universe. We consider the different expansion factors and obtain potentials of tachyonic scalar field. Inflection points of homogeneous tachyonic scalar field is calculated for these potentials, then we employed the tools of dynamical system analysis for the  considered  potentials and obtained the stable points.   
\end{abstract}

\section{Introduction}\label{sect.1}
 The cosmic acceleration is now an established phenomenon. As supported by several cosmological and astrophysical observations \cite{b1,b2,b3,b4,b5,b6,b7,b8,b9,b10} our universe is in the phase of accelerated expansion at the present stage and generally, it is believed that behind this accelerated expansion there is a mysterious component with negative pressure dominating $70\%$ of the universe. One of the good candidates that can explain this acceleration is ``Dark Energy" \cite{b11,bb11,b12,b13,b14} which is introduced as modified matter in Einstein's field equation which sits with the energy-momentum tensor. One of the remarkable features of dark energy is its negative pressure and repulsive gravity characteristics led by the equation of state $w_{de}<-1/3$. In alternative gravity theories, cosmic acceleration can be accounted for by modifications to the geometrical aspects of the field equation. Observationally,  the data from Planck-2018 CMB \cite{b7,b8,b9,b10} proposes that $70\%$ of the energy budget in the universe is in the form of dark energy while the remaining $30\%$ of the energy budget of the universe is attributed to non-relativistic baryonic and non-baryonic pressureless ( $p_m=0$) dust matter with equation of state(EoS) $w_{m}=0$. In recent years, various candidates have been put forth to elucidate the nature of dark energy, with the cosmological constant standing out as one of the simplest models, embodying constant dark energy with an EoS of $w_{\Lambda}= -1$.  Despite its alignment with observational data to a certain accuracy, the $\Lambda$CDM model grapples with unresolved cosmological constant and coincidence problems \cite{b15,b16,b17,b18,b19}. In response to these limitations, dynamical dark energy models, particularly those involving scalar fields, have gained prominence \cite{pn,p6,p7,p8,p9,p10}, with scalar field serving as a candidate of dynamical dark energy dubbed as $\phi$CDM having dynamical EoS $w_{\phi}$.\\
 
The nature of dark energy has been the subject of several literature; as alternatives to the cosmological constant, dynamic dark energy models have been developed, in which the EoS varies with time. There are many different dynamical dark energy models like quintessence field having tracker behaviour, late time accelerating solutions of modified Friedmann equations lead by modification of gravity, some other models include tachyon models, holographic dark energy models, chaplygin gas, phantom and k-essence \cite{s1,s2,ss1,s3,s4,s5,s6,s7}.\\
\\
%We consider the dynamical dark energy model $\phi$CDM in which quintessence \cite{pn} is deemed to be its simplest candidate and minimally coupled to matter \cite{L1}. Extensive research into the cosmological dynamics of different scalar fields  (Phantom and Tachyon) has been documented in the literature \cite{s1,s2,s3,s4,s5,s6,s7}.In cosmology, the theory of cosmic inflation (exponential expansion of the universe) is addressed by Guth and Linde independently  \cite{guth,AD} to overcome the Horizon and Flatness problem. Recently, the inflation model with the inflection point studied by several authors \cite{new1,new2,new3,new4,new5, new6, new7, new8, new9, new10, new11, new12, new13, new14, new15,,new16,new17,new18,new19,new20,new21}.
In our paper, we have focused on the tachyonic field as a candidate for dark energy. This type of scalar field is derived from string theory as the negative-mass mode of the open string perturbative spectrum. Though its primary use in the dark energy sector is phenomenological, we have investigated the behavior of the tachyonic scalar field near the inflection point. Recently, several authors have studied dark energy models with inflection points in the context of inflation. \cite{new1,new2,new3,new4,new5, new6, new7, new8, new9, new10, new11, new12, new13, new14, new15, New15,new16,new17,new18,new19,new20,new21}.\\
\\
\\
To obtain a contemporary value of EoS $w_\phi\approx-1$ which leads to accelerated expansion, the mechanism of slow rolling is a key ingredient. By imposing the approximation in which the kinetic term $\dot{\phi}^{2}/2$ of the tachyon is significantly smaller than unity i.e. $\dot{\phi}^{2}/2<< 1$ this can be achieved by positioning $\phi$ in an extremely flat region of the potential. Potentials with an inflection point are good candidates for providing flat regions suitable for the slow roll approximation. We consider the late-time behavior of the tachyon scalar field in the accelerated expansion of the universe, which shares similarities with early inflation. Numerous authors have examined the cosmic behavior of the tachyonic scalar field \cite{T1, TT1, T2}, discovering possibilities for both pure exponential growth and power-law expansion. We try to investigate the possibilities based on the superposition of these forms because the current data have not yet determined the precise shape of the evolution of the scale factor.

\textbf{\subsection{Dynamical set up }}\label{sec1.1}
In this section, we present the mathematical framework on which the dynamics are based. We choose GR to be the background theory. 
The cosmological principle is encoded in the FLRW metric when expressed in terms of spherical co-moving coordinates looks
\begin{equation} \label{e1}
 ds^2=g_{\mu\nu}dx^\mu dx^\nu\equiv-dt^2+a^2(t)\left(\frac{dr^2}{1-kr^2}+r^2 d\theta^2+r^2\sin^2\theta d\phi^2 \right).
\end{equation}
Here we have taken $c=1$. In the above equation, $a(t)$ is the scale factor and the parameter $k$ is the spatial curvature having values   $ k= -1,0,1 $ corresponding to spatially open, flat, and closed geometry respectively. For our model, we have taken the effective four-dimensional action for the tachyonic field which is given as \cite{A1,A2}:
\begin{equation}\label{e2}
 S=  \frac{1}{2\kappa^{2}}\int d^{4}x f(\phi)\sqrt{-g} R +  \int d^{4}x  V(\phi)\sqrt{-g}\sqrt{\left(1+g^{\mu\nu} \partial_\mu\phi\partial_\nu\phi\right)},
\end{equation}
where $V(\phi)$ is the potential of tachyonic scalar field and $\kappa^{2}=\frac{1}{M^{2}_{p}}=1$ set the 4D Planck scale considering $M_{p}$ is Planck mass as well as function of tachyon field $f(\phi)=1$ under minimal coupling. 
One can obtain the energy-momentum tensor for this action as
%----------------------------
\begin{equation}\label{e3}
T^{(\phi)}_{\mu\nu}=\frac{V(\phi)\partial_\mu\phi\partial_\nu\phi}{\sqrt{1+g^{\alpha\beta}\partial_\alpha\phi\partial_\beta\phi}}-g_{\mu\nu}V(\phi)\sqrt{1+g^{\alpha\beta}\partial_\alpha\phi\partial_\beta\phi}
\end{equation}
 %-----------------------------------
 Einstein's field equation for a spatially flat universe ($K=0$) gives the Friedmann equations as:
%----------------------------------------------------
\begin{equation}\label{e4}
  H^2=\left(\frac{\dot{a}}{a}\right)^2=\frac{\kappa^2}{3}\rho,
  \end{equation}
%----------------------------------------------------
  and
%----------------------------------------------------
  \begin{equation}\label{e5}
    \frac{\ddot{a}}{a}=-\frac{\kappa^2}{2}(\rho + p),
  \end{equation}
%----------------------------------------------------
 where $H(t)\equiv \frac{d a}{dt}$ is the Hubble parameter, $\rho$ and $p$ are the energy and the pressure respectively. In general, there could be multiple components  of matter contributing to energy-momentum tensor, 
 each of which with a different equation of state. However, in the post-recombination era (after $z\longrightarrow 1000)$ and more effectively towards the present epoch $(z\longrightarrow 0)$ matter and dark energy dominate the universe. We are therefore driven to focus on only two components: matter and dark energy. These two components fit together so perfectly that, if one is matter $w_m \approx 0$, the other is the dark energy $(w_{de} \approx -1$). For cosmic acceleration, we have $\rho + 3p < 0$.
The continuity equation is given as: 
%------------------------------------------------------------------------------------------------------------------------------------------------------------------------------------------------------
\begin{equation}\label{e6}
  \dot{\rho}+3H(1 + w)\rho=0.
\end{equation}
%----------------------------------------------------
%For constant EoS $w$ one can obtain from (\ref{e6}) the following scaling solution of $\rho$ for constant w as given by:
%\begin{equation}\label{e7}
 % \rho=\rho_0\left(\frac{a}{a_0}\right)^{-3(w+1)}
%\end{equation}
%where $\rho_{0}, a_{0}$ are present value of %$\rho $ and $a$ respectively.
%on can obtain the following form of expansion factor with the constant EoS $w$ as: 
%\begin{equation}\label{e8}
 % a(t)=a_0\left(\frac{t}{t_0}\right)^{\frac{2}{3(1+w)}}
  %\end{equation}
%where $t_{0}$ is the present epoch.
\\
In Sec.\ref{sec2} investigation of the behaviour of tachyonic scalar field is carried out near the inflection point. We have considered three different possibilities of cosmic behaviours; first two cases of conventional power-law expansion and pure exponential growth i.e. $a(t)=\alpha t^n, a(t)=\gamma e^{\beta t}$ respectively and the third case is of quasi-exponential expansion of the scale factor $a(t)=\eta t^n e^{\beta t}$ for these three cases we have calculated the respective potentials and then we investigated the behaviour of that particular model near their respective inflection points. In Sec.\ref{sec3} dynamical analysis for the above two cases is carried out. \\

%In the literature \cite{L2,L3} author investigates the behaviour of the quintessence near the inflection point by considering the cubic potential $V(\phi)=V_3(\phi-\phi_0)^3$. Under slow roll approximation ($\dot{\phi}^{2}/2<< V(\phi)$ ) quintessence leads to exponential expansion and transient acceleration near the inflection point.  Near the inflection point, one might refer to a moment or a particular set of conditions where the inflationary expansion changes behaviour.
%\section{$\phi$CDM Model: Background Mathematical Theory}\label{sec.0}

\section{Inflection point of tachyonic scalar field:} \label{sec2}
Using Eq.(\ref{e3}) for flat ($K=0$) FRLW universe the energy density and pressure for the tachyonic field are given by:
 %--------------------------
 \begin{equation}\label{e9}
\rho_\phi=\frac{V(\phi)}{\sqrt{1-\dot{\phi}^2}},
 \end{equation}
%---------------------------
\begin{equation}\label{e10}
    p_\phi=-V(\phi)\sqrt{1-\dot{\phi}^2}.
\end{equation}
%-----------------------------
Here dot signifies the derivative with respect to time. $V(\phi)$ is the potential associated with the tachyonic field that we have calculated in the upcoming subsections for each case. 
%--------------------------------------------------------------------------------
Now by using Eq.(\ref{e9}) and Eq.(\ref{e10}) the model may be written as fluid. By using:
\begin{equation}
    \frac{\Ddot{a}}{a}=H^2+\dot{H}^2
\end{equation}
and plugging Eq.(\ref{e9}) and Eq.(\ref{e10}) in Eq.(\ref{e4}) and (\ref{e5}) we get the following form of the second Friedmann equation:
\begin{equation}\label{e11}
    \frac{\Ddot{a}}{a}=\frac{\kappa^2 V(\phi)}{3\sqrt{1-\dot\phi^2}}\left(1-\frac{3}{2}\dot\phi^2\right)
\end{equation}
Now as we know for the accelerated expansion of the universe we have $\Ddot{a}>0$ thus using Eq.\ref{e11}; $\dot{\phi}^2 < \frac{2}{3}$  and the EoS (Equation of state) for $\phi$ is given by:
\begin{equation}\label{n11}
    w_\phi=\frac{p_\phi}{\rho_\phi}=(\dot\phi^2 - 1)
\end{equation}
For cosmic acceleration the possible range of the values of $w_\phi$ is $-1\leq w_\phi<-\frac{1}{3}$ and $0<\dot\phi^2<\frac{2}{3}$.\\
Using Eq.(\ref{e9}) and (\ref{e10}) one can easily find the equation of motion for this tachyonic model as:
\begin{equation}\label{ee12}
    \frac{\ddot{\phi}}{(1-\dot\phi^2)}+3H\dot\phi+\frac{1}{V}\frac{dV}{d\phi}=0.
\end{equation}
Now using Eq.(\ref{e4}) and Eq.(\ref{e5}); $\phi(t)$ and $V(\phi)$ can easily be written in the from of $H$ and $\dot{H}$ as following :
\begin{equation}\label{e13}
\phi(t)=\int{\left(dt \sqrt{-\frac{2\dot{H}}{3H^2}}\right)}
\end{equation}
\begin{equation}\label{e14}
 V(t)=\frac{3H^2}{\kappa^2} \sqrt{1+\frac{2\dot{H}}{3H^2}}
\end{equation}
Now using above two equations we will find the potential and their respective inflection point behaviour for the three cases to be discussed below.
\subsection{For Case-I: $a(t)=\alpha t^{n}$ }\label{sec2.1}
In this subsection we consider the simplest model for describing the expansion of the universe in certain cosmological scenarios, namely, the power law form of scale factor suggesting that the universe's expansion follows a simple power-law relationship with time. i.e. $a(t)=\alpha t^n$. Here $\alpha$ is a constant describing the overall scale of the universe and the exponent $n$ determines the rate of expansion and is greater than $0$ for the universe to experience the accelerated expansion. For this kind of model the Hubble parameter is of the following form:
\begin{equation}\label{e15}
    H=\frac{n}{t}
\end{equation}
Now on substituting this value of $H$ in Eq.(\ref{e14}) and using Eq.(\ref{e13}) we found the following form of the potential:
\begin{equation}\label{e16}
    V(\phi)=V_0 (\phi-\phi_0)^{-2}
\end{equation}
here $V_0 = \frac{2n}{\kappa^2}\sqrt{1-\frac{2}{3n}}$.
\begin{figure}[h!]
\centering
    \includegraphics[width=0.6\textwidth]{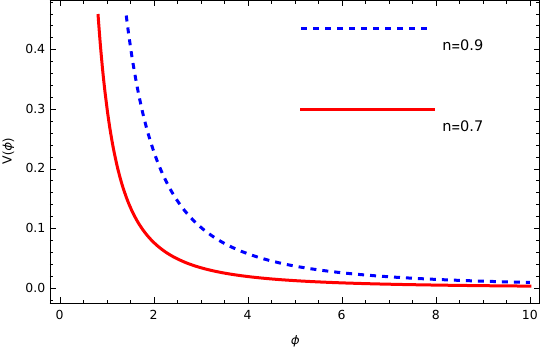}
    \caption{The plot shows the variation of $\phi$ with potential $V(\phi)$ for different values of the exponent factor $n$ ($n=0.7$ (red solid line) and $n=0.9$ (blue dotted line))}
    \label{fig:1(a)}
    \end{figure}
From above Eq.(\ref{e16}) we found there is no such inflection point for this particular model and we can also see that from the Fig.
(\ref{fig:1(a)}) as well that the curve show no inflection but one can observe that the scalar field is rolling slowly along the potential. Now using Eq.(\ref{ee12}) and on applying the slow roll approximation i.e. $\dot\phi^2<< 1 \implies 1-\dot\phi^2\approx 1$ we found the following evolution of scalar field with respect to time: 
\begin{equation}\label{e17}
    \phi(t)=\sqrt{\frac{2}{3n}(t^2-t_0^2)}+\phi_0
\end{equation}
here $t=t_0$ when $\phi=\phi_0$.\
\begin{figure}[h!]
\centering
    \includegraphics[width=0.7\textwidth]{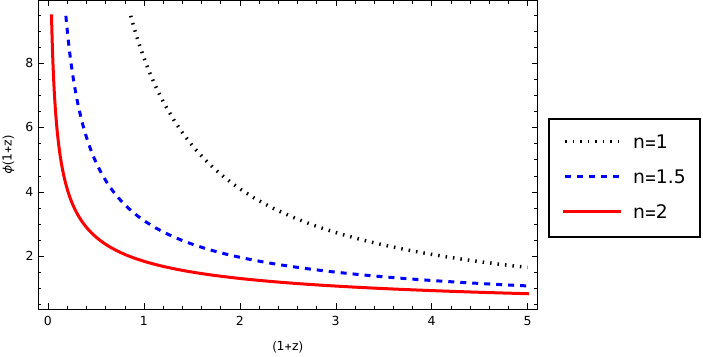}
    \caption{The plot shows the evolution of $\phi$ with cosmological redshift $(1+z)$ for different values of the exponent factor $n$; with $\alpha=0.1$ and $\phi_0<<1$.  }
    \label{fig:1(b)}
    \end{figure}
%--------explanation of evolution of phi-----
 In terms of cosmological redshift we can also rewrite the above Eq.(\ref{e17}) as:
\begin{equation}\label{ee17}
\phi(1+z)=\sqrt{\frac{2}{3n \alpha^{2/n}}} \sqrt{\left(\left(\frac{1}{1+z}\right)^{2/n}-\left(\frac{1}{1+z_0}\right)^{2/n}\right)}+\phi_0
\end{equation}
Fig.(\ref{fig:1(b)}) shows the evolution of $\phi$ for different values of $n$. We examine the behavior of a tachyonic scalar field, $\phi$, as the redshift, $(1+z)$, approaches zero, $\phi(1+z)$ approaches infinity  corresponding to the late-time evolution of the universe. The tachyonic field is characterized by a unique potential which is derived from the scale factor, this potential governs its dynamics and the corresponding impact on cosmic expansion. As the universe evolves towards the present era, the tachyonic field enters a phase where its kinetic energy becomes negligible, and it slowly rolls down its potential. This slow-roll regime results in the field behaving similarly to a cosmological constant, with a near-constant energy density. As a consequence, the universe continues to expand at an accelerating pace.\\
\\
\\
\subsection{For Case-II: \textbf{$a(t)=\gamma e^{\beta t}$}}\label{sec2.2}
For this particular case we have taken a simplest exponential form of the scale factor $a(t)=\gamma e^{\beta t}$ where $\gamma$ and $\beta$ are the constants. $\beta$ determines the overall rate at which the scale factor changes and can have two possibilities for the values of first is $\beta >0$ implies the exponential growth of the universe i.e. expanding universe over time and the second possibility is $\beta <0$ implies the exponential decay i.e. contracting universe. As we know during early epoch the universe experiences expansion so we can discard the second possibility. Now the second constant  $\gamma$ is a normalization factor that sets the overall scale of the universe at a particular time $t$ and $H$ for this exponentially expanding universe is proportional to the constant:
\begin{equation}\label{e18}
    H= \beta
\end{equation}
Now again using this value of $H$ and Eq.(\ref{e13}) we hot the following form of potential: 
\begin{equation}\label{e19}
    V(\phi)=\frac{3\beta^2}{\kappa^2}
\end{equation}
therefore for this particular model there is no inflection point because the potential is coming out to be constant and is consistent with the case that of quintessence model. From Eq.(\ref{ee12}), (\ref{e19}) and after applying the slow roll approximation the scalar field $\phi$ is coming out to be a constant and same result can be obtained from Eq.(\ref{e11}).

\subsection{For Case-III: \textbf{$a(t)=\eta t^n e^{\beta t}$}}\label{sec2.3}
\begin{figure}[h!]
\centering
    \includegraphics[width=0.7\textwidth]{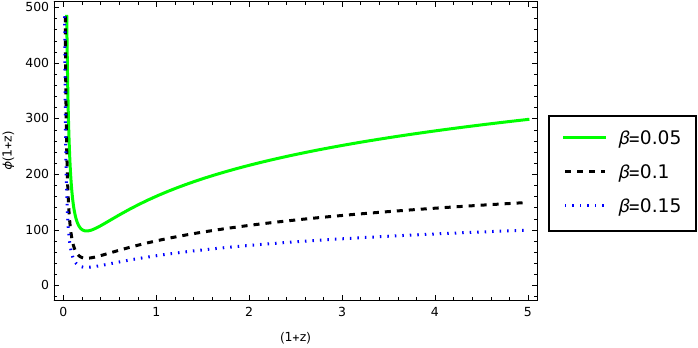}
    \caption{The plot shows the evolution of $\phi$ with cosmological redshift $(z+1)$ at inflection point with different values of the exponent factor $\beta$ for the simplest case when $n=1$ and $\eta=1$.}
    \label{fig:3(a)}
    \end{figure}

For this case we consider the universe the mixed form of the scale factor given by $a(t)=\eta t^n e^{\beta t}$ i.e. product of power law and exponential terms, here $\eta$, $n$ and $\beta$ are the constants determining overall rate at which the scale factor changes. For this model the Hubble parameter is obtained as:
\begin{equation}\label{e20}
    H=\frac{n}{t}+\beta
\end{equation}
Now using Eq.(\ref{e13}), (\ref{e14}) and (\ref{e20}), one can found the following form of potential and the scalar field in terms of $t$ as follow:
\begin{equation}\label{e21}
    V(t)=\frac{3(n+\beta t)^2}{\kappa^2 t^2}\sqrt{1 -\frac{2n}{3(n+\beta t)^2}}
    \end{equation}
and 
\begin{equation}\label{e22}
    t= -\frac{n}{\beta}+\frac{1}{\beta} \exp{\left(\beta\sqrt{\frac{3}{2n}}(\phi-\phi_0)\right)}
\end{equation}
On substituting  the Eq. (\ref{e22}) in (\ref{e21}) we got the following form of potential $V(\phi)$ which depends in the scalar field $\phi$ as:
%\begin{equation}\label{e23}   V(\phi)=\frac{2n}{\kappa^2 (\phi-\phi_0)^2} \sqrt{\left(1+\sqrt{\frac{3}{2n}}\beta (\phi-\phi_0)\right)^2-\frac{2}{3n}}\end{equation}
\begin{equation}\label{e23}
    V(\phi)=\frac{3\beta^2 \exp{\left(2\beta\sqrt{\frac{3}{2n}}(\phi-\phi_0)\right)}}{\left(\exp{(\beta\sqrt{\frac{3}{2n}}(\phi-\phi_0))}-n\right)^2} \sqrt{1-\frac{2n}{3\exp{\left(\beta\sqrt{\frac{3}{2n}}(\phi-\phi_0)\right)^2}}}
\end{equation}
%\begin{figure}[h!]
%\centering
 %   \includegraphics[width=0.6\textwidth]{case3potential.pdf}
%    \caption{The plot shows the evolution of potential $V''(\phi)$ with $\phi$ giving the inflection points for this particular model } \label{fig:3(a)}
%    \end{figure}
By using binomial expansion and on simplification, Eq.(\ref{e23}) can be rewritten as:
\begin{equation}\label{ee23}
    V(\phi)=3\beta^2 \left(\frac{\exp\left(2\beta\sqrt{\frac{3}{2n}}(\phi-\phi_0)\right)-n/3}{\left(\exp{\left(\beta\sqrt{\frac{3}{2n}}(\phi-\phi_0)\right)}-n\right)^2}\right)
\end{equation}
here for the simple case we use the following assumption:
\begin{equation}
\exp{\left(\beta\sqrt{\frac{3}{2n}}(\phi-\phi_0)\right)}\approx 1+\left(\beta\sqrt{\frac{3}{2n}}(\phi-\phi_0)\right)
\end{equation}
and solving above Eq.(\ref{ee23}) using the approximation mentioned above, one can easily find that the potential shows inflection at: 
\begin{equation}\label{ee24}
    \phi=\frac{2}{3}\sqrt{\frac{2n}{3\beta^2}}\left(\frac{(n-4)(2n-3)}{(n-3)(n-8)}\right)+\phi_0
    \end{equation}
For $n=3$ or $n=8$ inflection point approaches infinity and for the case when $n=4$ or $n=3/2$ inflection point becomes $\phi\longrightarrow\phi_0$. On substituting the value of potential (\ref{e23}) in equation of motion for this tachyonic field model, for the simplest case of $n=1$, we get the following form of the evolution of scalar field with respect to cosmological redshift at inflection point:
\begin{equation}\label{e24}
    \phi(1+z)=\frac{4}{\beta}\sqrt{\frac{3}{2}}\left(\frac{1}{4\eta(1+z)}-\log{\left(\frac{1}{4\eta(1+z
)}\right)}\right)
\end{equation}
Fig.(\ref{fig:3(a)}) shows the evolution of scalar field with respect to the redshift $(1+z)$. From the Fig.(\ref{fig:3(a)}) and Eq.(\ref{e24}) we can see that the scalar field evolves asymptotically as $(1+z)\longrightarrow 0$ signifying the phase of accelerated expansion. Using Eq.(\ref{e24}) and Eq.(\ref{n11}), the EoS (Equation of State) parameter of this inflection point tachyonic scalar field model $w_\phi\approx (0.4(1-4\eta(1+z))-1) \longrightarrow -0.8$ as $(1+z)\longrightarrow0$ with $\beta\approx0.05$ supporting eternal accelerated expansion of the universe at late time.
\\

  %  \begin{figure}[h!]
%\centering
 %   \includegraphics[width=0.7\textwidth]{fig4.pdf}
  %  \caption{The plot shows the evolution of $\phi$ with time $t$ for the second inflection point for the  different values of the exponent factor $n$ and $\beta$ }
   % \label{fig:3(b)}
    %\end{figure}
\section{Dynamical Analysis:}\label{sec3}
\begin{figure}[h!]
    \centering
    \includegraphics[width=0.6\textwidth]{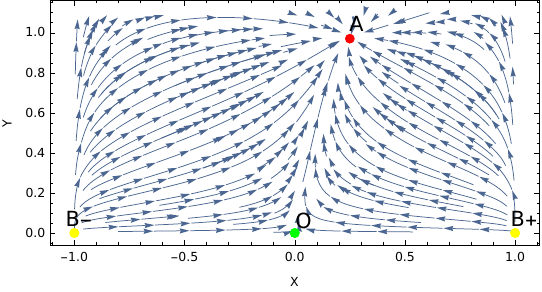}
    \caption{phase portait of the dynamical system with $\lambda\longrightarrow1$ and $w=0$ near inflection point. Here point $A$ in the phase space shows the region where universe undergoes accelerated expansion  }
    \label{fig:4a}
\end{figure}

In this section we performs the dynamical analysis of our inflection point tachyon model. Dynamical analysis is the study of the evolution of the universe, its behaviour over the time and allow us to analyse the various components of the universe like dark matter, dark energy, ordinary matter. The background equations are rewritten as an independent system of first order ODEs so that the stability at equilibrium points may be examined. Setting the dimensionless variables as:
\begin{subequations}
\begin{equation} \label{25a}
X=\dot{\phi} 
\end{equation}  
\begin{equation} \label{25b}
Y=\frac{\kappa\sqrt{V}}{\sqrt{3}H}
\end{equation}
\begin{equation} \label{25c}
\Gamma =\frac{V (d^2V/d\phi^2)}{(dV/d\phi)^2}
\end{equation}
\begin{equation} \label{25d}
    \lambda=-\frac{dV/d\phi}{\kappa\sqrt{V^3}}
\end{equation}
\end{subequations}
Here we have obtained an autonomous 2 dimensional system which has been studied by {\cite{T2}} and other two parameters (\eqref{25c}) and (\eqref{25d}) still has $\phi$ dependency and on differentiating above equations with respect to $\eta=loga$ we get the following set of 2 dimensional autonomous equations:
\begin{subequations}
    \begin{equation} \label{26a}
    X'=\frac{dX}{d\eta}=(X^2-1)(3X-\lambda\sqrt{3}Y)
\end{equation}
\begin{equation} \label{26b}
    Y'=\frac{dY}{d\eta}=\frac{-Y}{2}\left(\frac{3Y^2}{\sqrt{1-X^2}}(w_m-X^2+1)-3(w_m+1)+\sqrt{3}\lambda XY\right)
\end{equation}
\begin{equation}\label{26c}
    \lambda'=\frac{d\lambda}{d\eta}=-\left(\frac{\sqrt{27}}{2}-\sqrt{3} \Gamma\right)XY\lambda^2
\end{equation}
\end{subequations}
With a constraint equation:
\begin{equation}\label{e27}
    \frac{Y^2}{\sqrt{1-X^2}}+\Omega_m =1 
\end{equation}
The equation of state and the energy density of the tachyonic scale field in terms of these dimensionless parameters can be written as:
\begin{equation}
    w_\phi=X^2-1
\end{equation}
\begin{equation}
    \Omega_\phi=\frac{Y^2}{\sqrt{1-X^2}}
\end{equation}
and the effective equation of motion can be written as:
\begin{equation}\label{30}
    w_{eff}=w_m\left(1-\frac{Y^2}{\sqrt{1-X^2}}\right)-Y^2\sqrt{1-X^2} \\
\end{equation}
The critical points for the Eq.(\eqref{25a}) and (\eqref{25b}) are discussed below and are listed in the table(\ref{tab:1}):\\
\begin{table}[h!]
    \centering
    \begin{tabular}{|c|c|c|c|c|c|}
    \hline    
         $X$ & $Y$ & $w_{eff}$ & $w_\phi$& $\Omega_\phi$ &  Stability \\
         \hline
         $0$&$0$ &$w_m$ &$-1$ &$0$  &  Saddle point \\
         \hline
         $\pm1$& $0$& $0$& $0$&$-$ & Unstable Nodes\\
         \hline
         $1/5$& $\frac{\sqrt{3}}{5\lambda}$&$-$ & $\approx-0.96$&$\approx1$ & Stable node \\
         \hline
      
    \end{tabular}
    \caption{Above table shows the critical points of the system (\ref{25a})-(\ref{30}) with stability properties.}
    \label{tab:1}
\end{table}
\\
For the inflection point there are four critical points as shown in fig(\ref{fig:4a}) and are discussed below:\\

 \textbf{(a.) Point O}: For this critical point $X=0$ and $Y=0$; the  energy density of tachyonic scalar field is coming out to be $0$ signifying the matter dominated epoch. From the Friedmann constraint (\eqref{e27}) we can see that $\Omega_m=1$ and $w_{eff}=w_m$ again signifying the matter dominated epoch, thus this origin point of phase space refers to the saddle point having following eigenvalues:
\begin{equation} \nu_1=-3 <0 \end{equation} and \begin{equation}    \nu_2=\frac{3}{2}(w_m+1) >0 \end{equation}
This saddle point has an attractor behaviour towards X-axis as shown in fig(\ref{fig:4a}) and at this point the value of $w_\phi=-1$ imitating the cosmological constant behaviour.\\

%\begin{figure}[h!]
 %   \centering
  %  \includegraphics[width=0.6\textwidth]{DAIP2.pdf}
   % \caption{shows the phase portait for the above mentioned set of dynamical equations with $\lambda\longrightarrow1$ and $w=0$. }
    %\label{fig:4b}
%\end{figure}
\textbf{(b.) Point B$\_$ :} Here for this point we got the following critical points $X=-1$ and $Y=0$. At this point the tachyonic equation of state vanishes i.e. $w_\phi=0$. At this critical point we have following eigenvalues:
\begin{subequations}\label{sub33}

\begin{equation}
    \nu_1=6 >0
\end{equation}
and
\begin{equation}
    \nu_2=\frac{3}{2}(w_m+1) >0
\end{equation}
\end{subequations}
\\
signifying this critical point B$\_$ is an unstable node.\\

\textbf{(c.) Point B$_{+}$:} (X=1,Y=0) this point also represent unstable node having same eigenvalues as mentioned above in Eq.\ref{sub33}  \\

\textbf{(d.) Point A:} $(X=\frac{1}{5},Y=\frac{\sqrt{3}}{5\lambda})$ is the stable node having eigen values:
 \begin{equation}
        \nu_{1,2}=\frac{1}{2}\left(-4.8-1.9 w_m \pm \sqrt{(0.7-4.3w_m-3.9w_m^{2})}\right) < 0
    \end{equation}

%\begin{table}[htbp]
 %  \centering
  %  \begin{tabular}{|c|c|c|c|c|c|}
%\hline
%\multicolumn{6}{|c|}{\textbf{for case:I ($a(t)=\alpha t^n$)}} \\
%\hline
 %       $X$ & $Y$ & $w_{eff}$ & $w$ & $\Omega_\phi$ & $stability$  \\
%\hline   
 %       $0$ & $0 $  & $w$  & $-1$ &  0 & $-$\\ 
 
  %     $ \pm1 $ & $0$  & $0$ & $0$ & $-$ & $-$\\   

   %    $\frac{Y\lambda}{\sqrt{3}}$ & $Y $  & $-1+\frac{Y^2\lambda^2}{3}$ & $-1+\frac{Y^2\lambda^2}{3}$ & $1$ & $-$ \\   
%\hline
%\end{tabular}
%\end{table}
%here\begin{equation}
 %   \lambda=\frac{2}{\kappa \sqrt{V_0}}
 %   \end{equation} 

%\begin{equation}
 %   Y=\sqrt{\left(\frac{-\lambda^2\pm \sqrt{\lambda^4+36}}{6}\right)}
%\end{equation}    
%\subsection{Case:III}
Fig.(\ref{fig:4a}) shows the phase portait of the derived set of autonomous equations and the region near the point A depict the accelerated expansion of the universe. 

\section{Conclusion}

In this paper the evolution of minimally coupled Tachyonic scalar field at the inflection point has been investigated in an accelerating universe. The inflection point  signifies a transition between regions of stability and instability and inflection point provides a flat region suitable for the slow roll approximation. Near this point, the field's dynamics has been investigated for different forms of expansion factors. %Tachyonic scalar fields have been proposed as an inflationary model in the early universe.
We have mainly considered the three categories of scale (expansion) factor namely: (a) Power law expansion ($a(t)=\alpha t^n$), (b) Exponential form ($a(t)=\gamma e^{\beta t}$) and (c) Mixed form ($a(t)=\eta t^n e^{\beta t}$); for all these categories, the potentials and the corresponding possible inflection points have been calculated and the dynamical system of equations for the model has been obtained. Out of these three categories, mixed form shows inflection at $\phi$ given in Eq.(\ref{ee24}) and near this point under slow roll approximation the evolution of scalar field with cosmological redshift $(1+z)$ shows the accelerated expansion of the universe at late times. In section (\ref{sec3}) dynamical analysis carried out shows at critical point $A(1/5, \sqrt{3}/5\lambda)$ as shown in Fig.(\ref{fig:4a}) and (\ref{fig:4b}) represents the  stability of the derived dynamical system of equations and the region near inflection point which shows the accelerated expansion of the universe.

In addition to what is being done, it is important to study the inflection point of scalar field models in the presence of quantum gravity effects. In particular, scalar field dynamics plays a significant role in the post bounce scenario of loop quantum cosmology as it gives rise to inflationary solution \cite{m1,m2}. We leave it as our future pursuit to investigate how inflection points are effected due to quantum geometric correction for scalar fields. 
\section*{Acknowledgement}
The research by M.K. was carried out in Southern Federal University with financial support from the Ministry of Science and Higher Education of the Russian Federation (State contract GZ0110/23-10-IF).
\\


\begin{thebibliography}{99}

\bibitem{b1} A. G. Riess, et al., \emph{Astron. J.}, \textbf{116}, 1009(1998). 
\bibitem{b2} S. Perlmutter, et al., \emph{Astrophys. J.},  \textbf{517}, 565(1999). 
\bibitem{b3} D. Spergel, et al., \emph{Astrophys. J. S.}, \textbf{148}, 175(2003). 
\bibitem{b4} H. Seo and D. J. Eisenstein,  \emph{Astrophys. J.}, \textbf{598}, 720(2003).
\bibitem{b5} C. Blake and K. Glazebrook: \emph{Astrophys. J.}, \textbf{594}, 665(2003). 
\bibitem{b6}  E. Komatsu, et al., \emph{Astrophys. J. S.}, \textbf{192}, 18(2011).
\bibitem{b7} N. Aghanim, et al. [Planck Collaboration], Planck 2018 results VI. Cosmological parameters : \emph{A $\&$ A}, \textbf{641}, A6(2020).
\bibitem{b8} N. Aghanim, et al. [Planck Collaboration], Planck 2018 results I. Overview and the cosmological legacy of Planck : \emph{A $\&$ A}, \textbf{641}, A1(2020).
\bibitem{b9} R.A. Knop, et al., \emph{ Ap. J.}, \textbf{598}, 102 (2003).
\bibitem{b10} A.G. Riess, et al., \emph{Ap. J.}, \textbf{607}, 665 (2004).
\bibitem{b11} George F.R. Ellis et.al., \emph{JCAP}, \textbf{04}, 026 (2016).
\bibitem{bb11} T. Patil, S. Panda and M. Sharma, \emph{EPJ C}, \textbf{83}, 131(2023)
\bibitem{b12} Marek Szyd lowski, and Aleksander Stachowski, \emph{JCAP}, \textbf{10}, 066 (2015).
\bibitem{b13} Y. Bisabr and H. Salehi, \emph{Class. Quantum Grav.}, \textbf{19},2369 (2002).
\bibitem{b14} Y. Fujii and T. Nishioka,  \emph{Phys. Rev. D}, \textbf{42}, 361 (1990).
\bibitem{b15} Philip Bull, et al., \emph{Phys. Dark Universe}, \textbf{12}, 56(2016).
\bibitem{b16} L. Perivolaropoulos and F. Skara, \emph{New Astron. Rev.}, \textbf{95}, 101659(2022).
\bibitem{b17} Antonino Del Popolo and Morgan Le Delliou, \emph{Galaxies}, \textbf{5}, 17(2017).
\bibitem{b18} S. Weinberg, \emph{Rev. Mod. Phys.}, \textbf{61}, 1(1989).
\bibitem{b19} A. V. Astashenok and A. Del Popolo, \emph{Class. Quantum Gravity}, \textbf{29}, 085014(2012).
\bibitem{pn} Bharat Ratra and P. J. E. Peebles, \emph{Phys. Rev. D}, \textbf{37}, 3406 (1988).
\bibitem{p6} Celia Escamilla-Rivera and Antonio Najera, \emph{ JCAP}, \textbf{03}, 060(2022).
\bibitem{p7} G. B. Zhao, M. Raveri, L. Pogosian, et al., \emph{Nat Astron}, \textbf{1}, 627(2017).
\bibitem{p8}   R. Mainini et al., \emph{ApJ}, \textbf{599}, 24(2003).
\bibitem{p9} P. Brax, C. Burrage, C. Englert and  M. Spannowsky, \emph{Phys. Rev. D}, \textbf{94}, 084054(2016). 
\bibitem{p10} E. J. Copeland, M. Sami, and S. Tsujikawa, \emph{Int. J. Mod. Phys. D},  \textbf{15}, 1753(2006).
\bibitem{L1} L.Amendola, \emph{Phys.Rev.D}, \textbf{64}, 043511(2000).
\bibitem{s1} M.M. Verma and S. D. Pathak, \emph{Astrophys Space Sci}, \textbf{350}, 381(2014).
\bibitem{s2} M.M. Verma and S. D. Pathak, \emph{Int. J. Mod. Phys. D}, \textbf{23(9)}, 1450075(2014). 
\bibitem{ss1} S. Panda and M. Sharma, \emph{Astrophys. Space Sci.}, \textbf{361}, 1-8(2016)

\bibitem{s3} M.M. Verma and S. D. Pathak, \emph{Astrophys. Space Sci.}, \textbf{344}, 505(2013). 
\bibitem{s4} D. Rajeeb Kumar, S. D.Pathak and V. K. Ojha, \emph{Chinese Physics C}, \textbf{47(5)}, 055102(2023). 
\bibitem{s5} V. B. Johri,  \emph{Phys. Rev. D}, \textbf{70},  041303R(2004). 
\bibitem{s6} V. B. Johri,  \emph{Class. Quantum Gravity}, \textbf{19},  5959(2002). 
\bibitem{s7} V. B. Johri,  \emph{Phys. Rev. D}, \textbf{63},  103504(2001). 
%\bibitem{guth} Alan H. Guth, \emph{Phys. Rev. D}, \textbf{23},  347(1988). 
%\bibitem{AD} A.D. Linde, \emph{Phys. Lett. B},\textbf{108}, 389 (1982).
 \bibitem{new1} S. Choudhury and A. Mazumdar, \emph{Nucl. Phys. B}, \textbf{9}, 386(2014).
\bibitem{new2} R. Allahverdi, B. Dutta, A. Mazumdar ,\emph{Phys. Rev. D.},\textbf{78}, 063507(2008).
%_______dynamics of inflection point inflation___________
\bibitem{new3} Y. Bai and D. Stolarski  \emph{ JPAC}, \textbf{03}, 091(2021).
\bibitem{new4} S. Choudhury, A. Mazumdar, E. Pukartas, \emph{JHEP},  \textbf{4}, 1-27(2014)
\bibitem{new5} K. Enqvist and A. Mazumdar, \emph{Phys. Rep.}, \textbf{380}, 99-234(2003).
%--------------Dynamics of MSSM flat directions consisting of multiple scalar fields-------------%
\bibitem{new6} K. Enqvist, A. Jokinen and A. Mazumdar, \emph{JCAP}, \textbf{01}, 008(2004).
%------------Inflection point inflation within supersymmetry------------------------------
\bibitem{new7} K. Enqvist, A. Mazumdar and P. Stephens, \emph{JCAP}, \textbf{2010}, 020(2010).
%-------MSSM flat direction inflation: slow roll, stability, fine tunning and reheating------
\bibitem{new8} R. Allahverdi et al., \emph{JCAP}, \textbf{2007}, 019(2007).
%--------------------Particle physics models of inflation and curvaton scenarios--------
\bibitem{new9} A. Mazumdar and J. Rocher, \emph{Phys. Reps.}, \textbf{497}, 85-215(2011).
%---------Primordial blackholes and gravitational waves for an inflection-point model of inflation---------------------
\bibitem{new10} S. Choudhury and A. Mazumdar, \emph{Phys. Lett. B}, \textbf{733}, 270--275(2014).
%____Choudhury S, Mazumdar A. Reconstructing inflationary potential from BICEP2 and running of tensor modes___________________
\bibitem{new11} S. Choudhury and A. Mazumdar, \emph{ arXiv preprint}, \textbf{ arXiv:1403.5549}, (2014).
%______Gauge-invariant inflaton in the minimal supersymmetric standard model__________________
\bibitem{new12} R. Allahverdi et.al., \emph{Phys. Rev. Lett.}, \textbf{97}, 191304(2006).
%_________Allahverdi R, Dutta B, Mazumdar A. Attraction towards an inflection point inflation__________
\bibitem{new13} R. Allahverdi, B.Dutta and A. Mazumdar, \emph{Phys. Rev. D},\textbf{78}, 063507(2008).
%____Choi SM, Lee HM. Inflection point inflation and reheating. The European Physical Journal C. 2016 Jun;76:1-5.________
\bibitem{new14} S. M. Choi and H. M. Lee, \emph{EPJ C}, \textbf{76}, 1-15(2016).
%_______Dimopoulos K, Owen C, Racioppi A. Loop inflection-point inflation. Astroparticle Physics. 2018 Dec 1;103:16-20.______
\bibitem{new15} K. Dimopoulos, C. Owen and A. Racioppi, \emph{Astropart. Phys.}, \textbf{103}, 16-20(2018).
\bibitem{New15} J. Kaur, S. D. Pathak and M. Y. Khlopov, \emph{Astropart. Phys.}, \textbf{157}, 102926(2024).
\bibitem{new16} Laura Iacconi et al., \emph{JCAP}, \textbf{06}, 007, (2022).
\bibitem{new17} Jose J. Blanco-Pillado et al., \emph{JCAP}, \textbf{02}, 034, (2013).
\bibitem{new18}  K. Kefala et al.,     \emph{Phys. Rev. D}, \textbf{104}, 023506 (2021).
\bibitem{new19}   S. DavidStorm and Robert J.Scherrer,   \emph{Phys. Lett. B}, \textbf{829},  137126(2022).    \bibitem{new20} R. Cerezo and J.G. Rosa, \emph{J. High Energ. Phys.}, \textbf{2013}, 24 (2013).
\bibitem{new21} Shaun Hotchkiss et al.,\emph{JCAP}, \textbf{06}, 002 (2011).
%.......tachyon model...............
\bibitem{T1} V. Gorini et al., \emph{Phys. Rev. D.}, \textbf{69}, 123512(2004) 
\bibitem{TT1} A. Singh, H. K. Jassal and M. SHarma, \emph{JCAP}, \textbf{2020}, 008(2020)
\bibitem{T2} E. J. Copeland et al.,
 \emph{Phys. Rev. D.}, \textbf{71}, 043003 (2005)
\bibitem{L2} H. Y. Chang and R. J. Scherrer, \emph{Phys. Rev. D}, \textbf{88}, 083003 (2013).
\bibitem{L3}  S. D. Storm and R. J. Scherrer, \emph{Phys. Lett. B .}, \textbf{829}, 137126 (2022).
\bibitem{A1} Bamba, Kazuharu, et al., \emph{Astrophys.\& Space Sci.}, \textbf{342}, 155-228(2012). 
\bibitem{A2} Yun-Song Piao et al., \emph{Phys. Lett. B.}, \textbf{570}, 1(2003).
\bibitem{m1} M. Sharma et al., \emph{JCAP}, \textbf{2018}, 003(2018)
\bibitem{m2} M. Shahalam et 
al., \emph{Phys. Rev. D}, \textbf{96}, 123533(2017) 

\end{thebibliography}
\end{document}